\pdfoutput=1
\documentclass[10pt, conference]{IEEEtran}
\usepackage[left=0.625in,right=0.625in,top=0.75in,bottom=0.96in]{geometry}
\usepackage{amsmath,amsfonts,amssymb}
\usepackage{algorithm}
\usepackage{algpseudocode}
\usepackage{graphicx}
\usepackage{textcomp}
\usepackage{xcolor}
\usepackage{bm}
\usepackage{cite}
\usepackage{amsthm}
\usepackage{tabularx}
\usepackage{booktabs}
\usepackage{graphicx}
\usepackage{silence}
\usepackage{subcaption}

\begin{document}
\bstctlcite{BSTcontrol}

\title{HALO: Hierarchical Auction-assisted Learning for Offloading in SAGIN}
\author{
\IEEEauthorblockN{Xuli Cai, Poonam Lohan, Sachin Ravikant Trankatwar, and Burak Kantarci}
\IEEEauthorblockA{\textit{University of Ottawa, Ottawa, ON, Canada}\\
\{xcai049, ppoonam, sravikan, burak.kantarci\}@uottawa.ca}
}
\vspace{-0.3in}
\maketitle


\begin{abstract}
In this paper, we investigate delay-aware task offloading and resource scheduling in a three-tier space–air–ground integrated network (SAGIN) consisting of IoT devices, UAV edge nodes, and a high-altitude platform station (HAPS). We formulate joint task association and continuous resource control (including bandwidth, transmit power, and CPU frequency allocation) as a non-convex mixed-integer nonlinear programming (MINLP) problem, which is inherently NP-hard. To capture fine-grained system dynamics, we introduce a macro-micro slot model that tracks cumulative transmission and computation progress over time. Based on this model, we propose HALO, a hierarchical auction-assisted learning framework that combines auction-based task association with hierarchical Proximal Policy Optimization (HPPO) for resource allocation. Simulation results under different traffic loads show that HALO consistently outperforms representative deep reinforcement learning (DRL) baselines. In particular, HALO achieves an average improvement of 3.06 percentage points in task success rate over PPO (corresponding to a  $3.4$\% relative gain) and shows consistently greater robustness than DDPG and SAC, with relative improvements of $10.6$\% and $4.8$\%, respectively. These results highlight HALO's ability to maintain stable and efficient performance under varying traffic conditions, making it well-suited for delay-sensitive SAGIN environments.
\end{abstract}


\section{Introduction}
Space-Air-Ground Integrated Networks (SAGINs) are increasingly recognized as a promising architecture for latency-sensitive IoT services by integrating terrestrial devices, UAV edge nodes, and a high-altitude platform station (HAPS) into a unified computing-communication system \cite{wangComputationOffloadingMultiAgent2024, wuMultiHAPAssistedComputationOffloading2025, jiaHierarchicalAerialComputing2023}.  
With this multi-tier structure, tasks can be flexibly offloaded across heterogeneous nodes, improving service coverage and resilience. However, achieving efficient and delay-aware scheduling in SAGIN remains a challenging problem due to the tight coupling between task association, wireless transmission, and distributed computation under dynamic queue backlogs, and heterogeneous resource constraints.\looseness=-1

Recent studies have built strong foundations for SAGIN-enabled communication and computing. Existing works cover global transmission optimization and wide-area data collection \cite{fanGATOGlobalTransmission2025}, hierarchical HAPS-UAV aerial computing \cite{jiaHierarchicalAerialComputing2023}, multi-agent offloading in aerial-edge systems \cite{wangComputationOffloadingMultiAgent2024}, and multi-HAP-assisted offloading \cite{wuMultiHAPAssistedComputationOffloading2025}. Other efforts investigate joint offloading/association/resource allocation \cite{nabiJointOffloadingDecision2025}, perception-aware offloading \cite{waqasPerceptionAwareOffloadingCollaborative2025}, and task-prediction-driven edge offloading \cite{yangTaskPredictionBasedEdge2025}. At the communication and control layers, prior work also addresses UAV-assisted robustness \cite{maoUAVAssistedCommunicationsSAGINISAC2025}, hybrid FSO/RF coverage and deployment \cite{mashikoCombinedControlCoverage2025}, channel modeling and estimation \cite{mengChannelModelingEstimation2023}, RIS-enabled multifunctional architectures \cite{shenMultifunctionalRIsEnabledSAGIN2026}, outage/fairness analysis \cite{tanOutageProbabilityPerformance2025}, blockchain-supported trusted offloading \cite{tangBlockchainBasedTrustedTraffic2022}, and game-theoretic resource distribution \cite{zhangEvolutionaryGameResource2025}.
Despite these advances, three limitations remain from a delay-aware scheduling perspective. First, many methods rely on macro-scale delay abstractions (e.g., average-rate or static-capacity assumptions) that miss fine-grained service dynamics. Second, queue evolution is often modeled coarsely, weakening cross-interval carry-over under bursty arrivals. Third, joint optimization formulations are often hard to deploy due to mixed discrete-continuous coupling and computational complexity.\looseness=-1

To address these challenges, we formulate the delay-aware task offloading and resource scheduling problem in a three-tier SAGIN as a non-convex MINLP. We then propose HALO, a hierarchical auction-assisted learning framework that combines auction-based task association with HPPO-based resource allocation, supported by a macro–micro slot design for fine-grained system dynamics.
The main contributions of this work are summarized as follows:
\begin{itemize}
    \item We propose a macro–micro slot scheduling framework for three-tier SAGIN that models transmission and computation dynamics for accurate latency characterization and consistent queue evolution.
    \item We design a UAV-centric auction for task association, combining latency-aware utility and resource-aware bidding for adaptive task–UAV matching under dynamic network conditions.
    \item We propose HALO, a two-stage hierarchical learning framework coupling auction-based association with HPPO-based resource allocation, coordinating discrete decisions and continuous control across timescales.
\end{itemize}
We evaluate HALO against single-layer DRL baselines, including PPO, DDPG, and SAC, and against non-learning scheduling rules
. Simulation results, averaged over multiple traffic loads, demonstrate consistent performance gains in task success rate, with an average improvement of 3.06 percentage points ($+3.4$\% relative) over PPO. Moreover, HALO reduces task completion latency by $14$-$32$\%. The remainder of this paper is organized as follows. Section II presents the system model and problem formulation. Section III introduces the proposed HALO framework. Section IV provides simulation results under varying traffic conditions. Section V concludes the paper.

\begin{figure}[t]
    \centering
    \includegraphics[width=1\columnwidth]{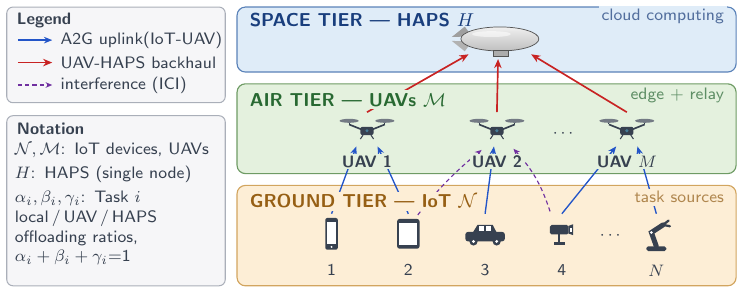}
    \caption{SAGIN System Model}
    \label{fig:system}
\end{figure}
\section{System Model and Problem Formulation}
\label{sec:model}

As illustrated in Fig.~\ref{fig:system}, we consider a three-tier SAGIN with ground IoT devices $\mathcal{N}=\{1,\dots,N\}$, UAVs $\mathcal{M}=\{1,\dots,M\}$ as aerial-edge nodes, and one HAPS as the high-altitude cloud tier. Each IoT device generates delay-sensitive tasks that may be processed locally, offloaded to an associated UAV, or further relayed to the HAPS. UAVs provide aerial edge computing and relay services, while the HAPS offers wider-area cloud-level computation.

\subsection{Task Arrival and Dynamic Queueing Model}
We adopt a mixed-timescale framework: macro-slots capture Poisson task arrivals with rate $\lambda$, while micro-slots enable fine-grained resource scheduling.  At each macro-slot, a task generated by IoT device $i$ is defined as:
\begin{equation}
\phi_i \triangleq (D_i, C_i, T_i^{\max}, t_i^{arr}),
\end{equation}
where $D_i$ denotes the data size, $C_i$ is the computation intensity, $T_i^{\max}$ is  latency constraint, and $t_i^{arr}$ is the arrival time. Each task is partitioned into local, UAV, and HAPS components with ratios $\alpha_i, \beta_i, \gamma_i$, respectively, satisfying $\alpha_i + \beta_i + \gamma_i = 1$.
The task partitioning variables are determined at macro slots by the Manager agent (detailed in Section~\ref{sec:method}). Whereas resource allocation is performed dynamically at each micro-slot rather than reserved over the entire task lifetime. Urgency-aware scheduling (e.g., based on remaining slack time) is applied to prioritize tasks. Completed transmission and computation stages are removed from the active scheduling set because they no longer require resources, allowing the released communication/computation capacity to be reallocated to unfinished tasks in subsequent micro-slots while preserving their accumulated progress and timing information.

\subsection{Channel Model}
\subsubsection{IoT-to-UAV Link ( Air-to-Ground (A2G) Channel)}
For IoT-UAV access links, blockage is modeled via probabilistic LoS/NLoS A2G propagation. The LoS probability between IoT $i$ and UAV $m$ is:
\begin{equation}
    P_{LoS}(\theta_{i,m}) = {1}/{[1 + a \exp \left( -b (\theta_{i,m} - a) \right)]},
\end{equation}
where $\theta_{i,m} = \frac{180}{\pi}\arcsin\!\left(\frac{H_m}{d_{i,m}}\right)$, $H_m$ is UAV altitude, $d_{i,m}$ is 3D distance between IoT device $i$ and UAV $m$, and $a,b$ are environment parameters. Then $P_{NLoS}=1-P_{LoS}(\theta_{i,m})$. Average path loss (dB) is:
\begin{equation}
    \bar{L}_{i,m} =
    P_{LoS}\left(L_{FS}+\eta_{LoS}\right)
    +P_{NLoS}\!\left(L_{FS}+\eta_{NLoS}\right),
\end{equation}
with $L_{FS}=20\log_{10}\!\left({(4\pi f_c d_{i,m})}/{c}\right)$, where $\eta_{LoS}$ and $\eta_{NLoS}$ denote the excess loss of the LoS and NLoS links, and $f_c$ and $c$ are the carrier frequency and the speed of light, respectively. The corresponding channel gain is
\begin{equation}\label{eq:a2g_gain}
    h_{i,m}=10^{-\bar{L}_{i,m}/10}.
\end{equation}

\subsubsection{UAV-to-HAPS Link (Air-to-Air(A2A) Channel)}
For UAV-HAPS backhaul, high-altitude propagation is LoS-dominant. The corresponding channel gain is:
\begin{equation}
    h_{m,H}=G_H\rho_0 d_{m,H}^{-\alpha_{AA}},
\end{equation}
where $\rho_0=(c/4\pi f_c)^2$ is the 1\,m reference gain, $G_H$ is the HAPS directional antenna gain, $d_{m,H}$ is UAV-HAPS distance, and $\alpha_{AA}\approx2$ is path loss exponent.

\subsection{Communication Model}
Let $T$ be macro-slot duration (index $t\in\{1,2,\dots\}$), and let each macro-slot contain $K$ micro-slots of duration $\Delta t$ (index $k\in\{1,2,\dots,K\}$) such that $T = K\Delta t$.
\subsubsection{IoT-to-UAV Uplink (Access Link)}
If task $i$ uses bandwidth fraction $w_i^{U}[t,k]$ and power $p_i^{IoT}[t,k]$ at micro-slot $k$, its uplink rate to UAV $m$ is:
\begin{equation}\label{eq:rate_uav}
\begin{aligned}
    R_{i,m}^{U}[t,k] &= w_i^{U}[t,k]B^{U}\\
    &\quad\times \log_2\!\left(1+\frac{p_i^{IoT}[t,k]h_{i,m}}{N_0 w_i^{U}[t,k]B^{U}+I_{i,m}[t,k]}\right),
\end{aligned}
\end{equation}
 where $B^{U}$ is the total bandwidth available at each UAV, $N_0$ is the AWGN power spectral density (PSD) and $I_{i,m}[t,k]$ denotes inter-cell interference (ICI) on task $i$.

Intra-cell FDMA eliminates intra-cell interference; only ICI is considered. For victim task $i$ (UAV $m$) and interferer $j$ (UAV $v\neq m$), define $
B_i[t,k]=w_i^{U}[t,k]B^{U}$, and
$B_j[t,k]=w_j^{U}[t,k]B^{U}$.
The interference PSD from $j$ at UAV $m$ is
\begin{equation}\label{eq:PSD}
    \rho_{j,m}[t,k]=\frac{P_{j,m}^{rx}[t,k]}{B_j[t,k]}=\frac{p_j^{IoT}[t,k]h_{j,m}}{w_j^{U}[t,k]B^{U}},
\end{equation}
where $h_{j,m}$ is the A2G gain of the interfering link, obtained from \eqref{eq:a2g_gain}.
With independent FDMA across cells, expected overlap bandwidth is
\begin{equation}\label{eq:Overlap}
    \mathbb{E}\!\left[B_{i,j}^{overlap}[t,k]\right]=\big(w_i^{U}[t,k]w_j^{U}[t,k]\big)B^{U}.
\end{equation}
Using Eq. \ref{eq:PSD} and Eq. \ref{eq:Overlap}, the ICI from task $j$ to task $i$ is:
\begin{equation}
\begin{aligned}
    I_{j\to i}[t,k]
    &=\rho_{j,m}[t,k]\cdot \mathbb{E}\!\left[B_{i,j}^{overlap}[t,k]\right]\\
    &=w_i^{U}[t,k] p_j^{IoT}[t,k]h_{j,m}.
\end{aligned}
\end{equation}
Thus, the total interference on task $i$ is
\begin{equation}\label{eq:interference}
    I_{i,m}[t,k]=w_i^{U}[t,k]\sum_{v\in\mathcal{M}\setminus\{m\}}\sum_{j\in\mathcal{N}_v[t,k]} p_j^{IoT}[t,k]h_{j,m},
\end{equation}
where $\mathcal{N}_v[t,k]$ is the active tasks set associated with UAV $v$, determined by the auction mechanism described in Section~\ref{sec:method}.
For offloaded data $(\beta_i+\gamma_i)D_i$, let the generation time coordinate be $(t_i^{arr},0)$ and the uplink completion time coordinates be $(t_i^{fin,uav,tx},k_i^{fin,uav,tx})$. The transmission duration is:
{\small
\begin{equation}
    T_{i,tx}^{UAV} = (t_i^{fin,uav,tx}-t_i^{arr}) T + k_i^{fin,uav,tx}\Delta t.
\end{equation}}
By aggregating the uplink rate over all active micro-slots within this interval, the completion condition is succinctly given by:
{\small
\begin{equation}
    \sum_{t=t_i^{arr}}^{t_i^{fin,uav,tx}} \sum_{k} R_{i,m}^{U}[t,k]\Delta t \ge (\beta_i+\gamma_i)D_i,
\end{equation}}
where the inner sum runs up to $K$ for intermediate macro-slots, and up to $k_i^{fin,uav,tx}$ in the final macro-slot.

\subsubsection{UAV-to-HAPS Uplink (Backhaul Link)}
At micro-slot $k$, UAV $m$ allocates $w_i^{H}[t,k]$ and $p_{i,m}^{UAV}[t,k]$ for relayed task $i$. Neglecting inter-UAV backhaul interference, the rate is:
{\small
\begin{equation}
    R_{i,m}^{H}[t,k] = w_i^{H}[t,k]B^{H}
    \times \log_2\left(1+\frac{p_{i,m}^{UAV}[t,k]h_{m,H}}{N_0 w_i^{H}[t,k]B^{H}}\right),
\end{equation}}
where $B^H$ denotes the total bandwidth available at HAPS.
For the relayed data $\gamma_iD_i$, backhaul starts and ends at $(t_i^{fin,uav,tx},k_i^{fin,uav,tx})$ and $(t_i^{fin,haps,tx},k_i^{fin,haps,tx})$, respectively. The total backhaul duration $T_{i,tx}^{HAPS}$ is calculated as the exact time elapsed between these two boundary coordinates. The capacity constraint over this period is:
{\small
\begin{equation}\label{eq:bh_constraint}
\sum_{t=t_i^{fin,uav,tx}}^{t_i^{fin,haps,tx}} \sum_{k} R_{i,m}^{H}[t,k]\Delta t \ge \gamma_iD_i.
\end{equation}}
\subsection{Computation Model}
CPU frequencies are scheduled at the micro-slot level. The total computation workload of task $i$ is $D_iC_i$, which is split into local, UAV, and HAPS parts according to $\alpha_i, \beta_i$, and $\gamma_i$. Similar to the transmission model, the exact computation duration for each tier is calculated as the time elapsed between its respective start and end coordinates.

\subsubsection{Local IoT Computing}
For the local portion, processing starts at $(t_i^{arr},0)$ and ends at $(t_i^{fin,loc,comp},k_i^{fin,loc,comp})$, yielding the local computation duration $T_{i,comp}^{loc}$. By aggregating the local CPU frequency $f_i^{loc}[t,k]$ over this period, the processed workload constraint is:
{\small
\begin{equation}
    \sum_{t=t_i^{arr}}^{t_i^{fin,loc,comp}} \sum_{k} f_i^{loc}[t,k]\Delta t \ge \alpha_iD_iC_i.
\end{equation}}
\subsubsection{UAV Edge Computing}
Edge processing can only begin after the A2G transmission is completed. It starts at $(t_i^{fin,uav,tx},k_i^{fin,uav,tx})$ and ends at $(t_i^{fin,uav,comp},k_i^{fin,uav,comp})$, taking a duration of $T_{i,comp}^{UAV}$. The allocated UAV CPU frequency $f_i^{U}[t,k]$ must satisfy:
{\small
\begin{equation}
\sum_{t=t_i^{fin,uav,tx}}^{t_i^{fin,uav,comp}} \sum_{k} f_i^{U}[t,k]\Delta t \ge \beta_iD_iC_i.
\end{equation}}
\subsubsection{HAPS Cloud Computing}
Similarly, HAPS processing starts after the A2A backhaul is completed and ends at $(t_i^{fin,haps,tx},k_i^{fin,haps,tx})$ and  $(t_i^{fin,haps,comp},k_i^{fin,haps,comp})$, respectively, with a duration of $T_{i,comp}^{HAPS}$. 
The required cloud workload constraint is:
{\small
\begin{equation}
\sum_{t=t_i^{fin,haps,tx}}^{t_i^{fin,haps,comp}} \sum_{k} f_i^{H}[t,k]\Delta t \ge \gamma_iD_iC_i.
\end{equation}}

\subsection{Total Delay Definition}
Under the partial offloading scheme, the local, edge, and cloud portions of a task are processed independently by separate computing resources after their required transmission stages are completed. Therefore, these branches can execute in parallel. Queueing delay is inherently captured by the scheduled micro-slot boundaries. When a task waits, it receives no resources and makes no progress, shifting its completion to a later micro-slot. Thus, the measured elapsed time already includes both waiting and service delays, avoiding the need for a separate queueing term.
The edge and cloud branches share the initial A2G transmission phase $T_{i,tx}^{UAV}$ and then diverge.
The overall task makespan is defined as the maximum latency among the three branches:
\begin{equation}
    T_i^{total}=\max\big\{T_i^{Local},\ T_i^{Edge},\ T_i^{Cloud}\big\},
\end{equation}
where the latency of each individual branch is given by: $T_i^{Local}= T_{i,comp}^{loc}$, 
    $T_i^{Edge} = T_{i,tx}^{UAV}+T_{i,comp}^{UAV}$, and
    $T_i^{Cloud} = T_{i,tx}^{UAV}+T_{i,tx}^{HAPS}+T_{i,comp}^{HAPS}$.
\begin{figure*}[t]
    \centering
    \includegraphics[width=0.9\textwidth]{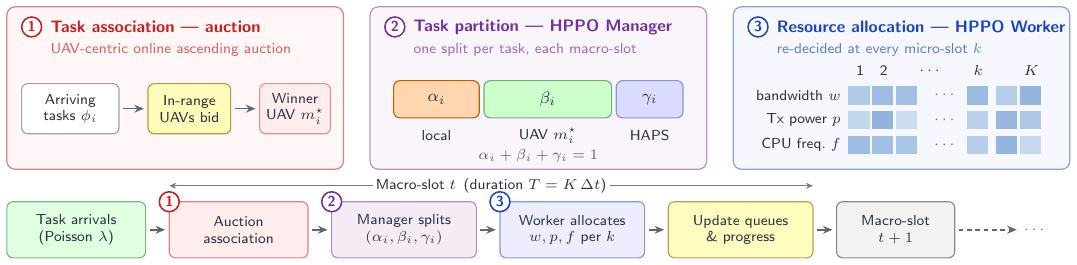}
    \caption{Proposed Framework: Task Offloading and Execution Flow (Macro - Micro Slot Operations)}
    \label{fig:proposal}
\end{figure*}
\subsection{Problem Formulation}
Define
$\mathbf{A}=\{\alpha_i,\beta_i,\gamma_i\}$,
$\mathbf{F}=\{f_i^{loc}[t,k],f_i^{U}[t,k],f_i^{H}[t,k]\}$,
$\mathbf{W}=\{w_i^U[t,k],w_i^H[t,k]\}$,
$\mathbf{P}=\{p_i^{IoT}[t,k],p_{i,m}^{UAV}[t,k]\}$.
The optimization problem is:
\begin{subequations}\label{eq:opt_prob}\small
\begin{align}
    &\max_{\mathbf{A},\mathbf{W},\mathbf{P},\mathbf{F}}\;\frac{1}{|\mathcal{N}_{all}|}\sum_{i\in\mathcal{N}_{all}}\mathbb{I}\!\left(T_i^{total}\le T_i^{\max}\right)\label{eq:obj}\\  &\text{s.t.}\quad\alpha_i+\beta_i+\gamma_i=1,\;\alpha_i,\beta_i,\gamma_i\in[0,1],\;\forall i,\label{eq:cons_ratio}\\
    &\quad\sum_{i\in\mathcal{N}_m^{tx}[t,k]}w_i^U[t,k]\le1,\;\forall m,t,k,\label{eq:cons_bw_uav}\\
    &\quad\sum_{i\in\mathcal{M}_H^{tx}[t,k]}w_i^H[t,k]\le1,\;\forall t,k,\label{eq:cons_bw_haps}\\
    &\quad p_i^{IoT}[t,k]\le P_{\max}^{IoT},\;\forall i,t,k,\label{eq:cons_pwr_iot}\\
    &\quad\sum_{i\in\mathcal{N}_m[t,k]}p_{i,m}^{UAV}[t,k]\le P_{\max}^{UAV},\;\forall m,t,k,\label{eq:cons_pwr_uav}\\
    &\quad f_i^{loc}[t,k]\le F_{\max}^{IoT},\;\forall i,t,k,\label{eq:cons_cpu_iot}\\
    &\quad\sum_{i\in\mathcal{N}_m[t,k]}f_i^{U}[t,k]\le F_{\max}^{UAV},\;\forall m,t,k,\label{eq:cons_cpu_uav}\\
    &\quad\sum_{i}f_i^{H}[t,k]\le F_{\max}^{HAPS},\;\forall t,k.\label{eq:cons_cpu_haps}
\end{align}
\end{subequations}

Problem~\eqref{eq:opt_prob} is a non-convex MINLP, since the task-UAV association is inherently discrete and is implicitly embedded in the association-dependent active sets $\mathcal{N}_m[t,k]$. The task success rate indicator objective and coupling among $\mathbf{A},\mathbf{W},\mathbf{P},\mathbf{F}$ induce bilinear and quadratic-fractional SINR terms, making the problem NP-hard. In addition, active sets (e.g., $\mathcal{N}_m[t,k]$) vary stochastically with arrivals/departures, so static convex methods are not directly applicable.\looseness=-1

\section{Proposed Solution Framework}
\label{sec:method}
To address the formulated problem, we propose \textbf{HALO} presented in Fig. \ref{fig:proposal}, a hierarchical auction-assisted learning framework for joint task association and resource allocation.

\subsection{UAV-Centric Online Auction Association}
\label{subsec:uav_online_auction}

Building on the combinatorial auction of \cite{hoosSolvingCombinatorialAuctions}, we design an online auction for UAV association. For each arriving task $i$ defined by the tuple $\phi_i = (D_i, C_i, T_i^{\max}, t_i^{arr})$, only UAVs within communication coverage are eligible bidders. The bidder set is strictly:
\begin{equation}
\mathcal{M}_i=\{m\in\mathcal{M}\mid d_{i,m}\le R_m\},
\end{equation}
where $R_m$ is the coverage radius of UAV $m$. UAVs outside coverage are never considered for bidding or assignment.

The task value reflects urgency and is defined as $v_i=D_i/T_i^{\max}$. The association cost accounts for task size and propagation distance: $
c_{i,m}=D_i\left(1+{d_{i,m}}/{d_0}\right),\quad m\in\mathcal{M}_i$.
where $d_0$ is a reference distance used to normalize the distance effect. Thus, larger tasks and farther UAVs incur higher association cost.
We apply batch-wise min--max normalization:

$\tilde v_i=\mathrm{Norm}(v_i)$,\qquad
$\tilde c_{i,m}=\mathrm{Norm}(c_{i,m})$,
and define the bid utility $
\pi_{i,m}=\tilde v_i-\tilde c_{i,m}$.
Tasks of the same arrival batch are cleared sequentially, each awarded to its best bidder,
\begin{equation}
m_i^\star=\arg\max_{m\in\mathcal{B}_i}\pi_{i,m},
\end{equation}
where $\mathcal{B}_i\subseteq\mathcal{M}_i$ is the capacity-aware bidder set. Specifically, $\mathcal{B}_i$ retains the UAVs of $\mathcal{M}_i$ that currently serve fewer than $N_{\mathrm{cap}}$ concurrent tasks; if every in-coverage UAV is at that cap, all of $\mathcal{M}_i$ bids.

With $\pi_i^{(1)}$ and $\pi_i^{(2)}$ the best and second-best utilities, task $i$ clears at the runner-up margin
\begin{equation}
b_i=\big(\pi_i^{(1)}-\pi_i^{(2)}\big)+\varepsilon,
\end{equation}
where $\varepsilon>0$ keeps the bid strictly positive under ties. One round settles each task because contention is carried by $\mathcal{B}_i$, which drops a UAV as soon as it reaches $N_{\mathrm{cap}}$ awards within the batch, so no price iteration is required.
Task $i$ is assigned to $m_i^\star$. The set of tasks associated with UAV $m$ at time $(t,k)$ is denoted by $\mathcal{N}_m[t,k]$, which forms the active association set used in the communication model. If $\mathcal{M}_i=\varnothing$, the task is not admitted and is excluded from \eqref{eq:obj}.
\begin{algorithm}[t]\footnotesize
\caption{{HALO (HPPO) Training}}
\label{alg:mlp-hppo-short}
\begin{algorithmic}[1]
\Require load set $\mathcal{L}$, episodes $E$, horizon $T$, worker update threshold $U_w$, manager update threshold $U_m$
\State Initialize manager/worker PPO agents and replay buffers $\mathcal{D}_m,\mathcal{D}_w$
\Comment{$U_w,U_m$ are minimum numbers of stored transitions before one PPO update}
\For{$e=1$ to $E$}
    \State Sample load $\ell\sim\mathcal{L}$ and reset environment
    \State Decay manager/worker learning rates and entropy coefficients
    \For{$t=1$ to $T$}
        \If{new tasks exist}
            \State Manager samples high-level action $a_m^t$
        \Else
            \State Use manager no-op action
        \EndIf
        \State Append one manager transition to $\mathcal{D}_m$
        \State Worker samples per-UAV low-level actions and executes low-level step(s)
        \State Append all valid worker transitions in this second to $\mathcal{D}_w$
        \Comment{$|\mathcal{D}_w|$ grows by the number of active UAV-task decisions}
        \State Compute manager reward and update the latest manager transition reward
    \EndFor
    \State Mark episode-end terminals in $\mathcal{D}_m,\mathcal{D}_w$
    \If{$|\mathcal{D}_w| \ge U_w$}
        \State Update worker by PPO-Clip+GAE; clear $\mathcal{D}_w$
    \EndIf
    \If{$|\mathcal{D}_m| \ge U_m$}
        \State Update manager by PPO-Clip+GAE; clear $\mathcal{D}_m$
    \EndIf
\EndFor
\State Save best manager/worker checkpoints
\end{algorithmic}
\end{algorithm}

\subsection{Two-Timescale Hierarchical PPO Architecture}
To prevent action-space explosion and non-stationarity caused by differing timescales, we adopt a hierarchical Manager-Worker DRL architecture.

\subsubsection{Stable Policy Optimization via PPO}
Both Actor-Critic agents use PPO to ensure stable updates and avoid congestion. Let $r(\theta)=\frac{\pi_\theta(a|s)}{\pi_{\theta_{\mathrm{old}}}(a|s)}$ be policy ratio. The clipped  objective is:
{\small
\begin{equation}
L^{\mathrm{CLIP}}(\theta)=\hat{\mathbb{E}}
\left[
\min\!\left(
r(\theta)\hat{A},\;
\mathrm{clip}(r(\theta),1-\epsilon,1+\epsilon)\hat{A}
\right)
\right],
\end{equation}}
where $\epsilon$ is the clipping coefficient and $\hat{A}$ is the GAE-estimated advantage. PPO ensures conservative policy evolution via clipping and cross-timescale variance reduction via GAE.

\subsubsection{Macro-Level Manager Agent (Task Splitting)}
At macro-slot $t$ (e.g., 1 s), the Manager assigns task splitting for arrivals.

\begin{itemize}
    \item \textbf{State Space ($\mathbf{s}^M_t$):} Processing up to $N_{\max}$ tasks (with zero-padding or truncation), each task $i$ is encoded as
    $$ \big[\tilde{d}_i,\;\tilde{\ell}_i,\;\tilde{q}^{\mathrm{uav}}_i,\;\tilde{q}^{\mathrm{haps}},\;\tilde{u}_i\big], $$
    capturing normalized size, distance, UAV/HAPS backlogs, and urgency. Concatenation yields
    $ \mathbf{s}^M_t \in \mathbb{R}^{5N_{\max}}$.

    \item \textbf{Action Space ($\mathbf{a}^M_t$):} The Manager outputs
    $\mathbf{z}^M_t \in \mathbb{R}^{3N_{\max}}$.
    For each task $i$, a 3-D slice is Softmax-normalized for local/edge/cloud splitting:
    \begin{equation}
    \mathbf{a}_{i,t}^{M}=
    \begin{bmatrix}
    \alpha_i\; \beta_i \;\gamma_i
    \end{bmatrix}^{\top}
=\mathrm{Softmax}\!\left(\mathbf{z}_{i,t}^{M}\right).
    \end{equation}

    \item \textbf{Reward Function ($r_t^M$):} Evaluated after all micro-slots from the on-time ($N_{\mathrm{succ}}$) and failed ($N_{\mathrm{fail}}$) completions, normalized by the arrival count $A_t$:
    \begin{equation}
    r_t^M=\big(\rho_{\mathrm{succ}}N_{\mathrm{succ}}-\rho_{\mathrm{fail}}N_{\mathrm{fail}}\big)/\max(1,A_t).
    \end{equation}
\end{itemize}

\subsubsection{Micro-Level Worker Agent (Resource Scheduling)}
At micro-slot $k$ (e.g., 50 ms), the Worker executes fine-grained resource allocation.

\begin{itemize}
    \item \textbf{State Space ($\mathbf{s}^W_{k,j}$):} Each resource queue $j$ is encoded as a Top-$K_w$ vector (sorted by least-slack-time):
    $ \mathbf{s}^W_{k,j}\in\mathbb{R}^{2K_w}$,
    comprising $K_w$ pairs of normalized residual workload and urgency
    $[\tilde{L}_{m},\tilde{U}_{m}]$.

    \item \textbf{Action Space ($\mathbf{a}^W_{k,j}$):} An MLP outputs
    $\mathbf{a}^W_{k,j}\in\mathbb{R}^{2K_w}$.
    For tx, the segments represent bandwidth and power logits:
    \begin{equation}
    \mathbf{a}_{k,j}^{\mathrm{tx}}=
    \begin{bmatrix}
    \mathrm{Softmax}(\mathbf{z}^{B}_{k,j})\;
    \sigma(\mathbf{z}^{P}_{k,j})
    \end{bmatrix}^{\top}.
    \end{equation}
    To satisfy hardware limits, requested powers sharing a transmitter are rescaled:
    \begin{equation}
    \tilde{p}_{e}[k]=
    \frac{p^{\mathrm{req}}_{e}[k]}
    {\max\!\left(1,\sum_{j\in\mathcal{E}^{\mathrm{tx}}_{n}[k]}p^{\mathrm{req}}_{j}[k]\right)}.
    \end{equation}
    Computation queues apply Softmax directly for CPU-share weights.

    \item \textbf{Reward Function ($r_{t,k}^W$):}
    The worker adopts a dense micro-step progress reward:
    \begin{equation}
    r_{t,k}^W
    = \sum_{j\in\mathcal{A}_{\mathrm{tx}}(t,k)} \Delta b_j(t,k)
    + \sum_{j\in\mathcal{A}_{\mathrm{comp}}(t,k)} \Delta c_j(t,k).
    \end{equation}
    \begin{equation}
    \Delta b_j(t,k)=\min\!\left\{R_j(t,k)\Delta t,\; b_j^{\mathrm{rem}}(t,k)\right\},
    \end{equation}
    \begin{equation}
    \Delta c_j(t,k)=\min\!\left\{f_j(t,k)\Delta t,\; c_j^{\mathrm{rem}}(t,k)\right\}.
    \end{equation}
\end{itemize}
The training process of the hierarchical PPO framework is summarized in Algorithm~\ref{alg:mlp-hppo-short}.

\section{Simulation Results}
\label{sec:results}
\subsection{Simulation Setup}

We consider 100 IoT devices distributed over a $0.9 \times 0.7$ km area using four Gaussian hotspots, with a standard deviation of approximately 78 m. Four UAVs are deployed using $k$-means clustering, while the HAPS is located above the center of the considered area. Among the 100 IoT devices, 98 are covered by at least one UAV. Specifically, 65 devices are covered by one UAV and 33 devices are covered by two UAVs, resulting in an average of 1.34 eligible UAV bidders per admitted task. The remaining two devices are outside the coverage of all UAVs, and their generated tasks are therefore dropped for all methods.

\begin{table}[t]
    \centering
    \caption{Simulation Parameters}
    \label{tab:sim_parameters}
    \footnotesize
    \setlength{\tabcolsep}{5pt}
    \renewcommand{\arraystretch}{0.92}
    \resizebox{\columnwidth}{!}{%
    \begin{tabular}{lcc}
        \toprule
        \textbf{Parameter} & \textbf{Symbol} & \textbf{Value} \\
        \midrule
        \multicolumn{3}{c}{\textit{Network \& Physical Parameters}} \\
        \midrule
        Number of IoT devices/UAVs & $N$ / $M$ & 100 / 4 \\
        UAV slant coverage radius & $R_m$ & 250 m \\
        UAV / HAPS altitude & - & 100 m / 20 Km \\
        Carrier frequency & $f_c$ & 2.0 GHz \\
        Noise PSD & $N_0$ & $-130$ dBm/Hz \\
        HAPS antenna gain & $G_H$ & 40 dBi \\
        A2G params / LoS,NLoS loss & $a,b$ / $\eta_{LoS}$,$\eta_{NLoS}$ & 9.61, 0.16 / 1, 20 dB \\
        Bandwidth IoT$\to$UAV / UAV$\to$HAPS & $B^{U}$ / $B^{H}$ & 20 / 100 MHz \\
        Max transmit power IoT / UAV & $P_{\max}^{IoT}$ / $P_{\max}^{UAV}$ & 0.5 / 2.0 W \\
        Max CPU freq.\ IoT / UAV / HAPS & $F_{\max}^{IoT}$/$F_{\max}^{UAV}$/$F_{\max}^{HAPS}$ & 1 / 4 / 30 GHz \\
        \midrule
        \multicolumn{3}{c}{\textit{Task \& Time Parameters}} \\
        \midrule
        Evaluation arrival rate (Poisson) & $\lambda$ & 300/350/400 tasks/min \\
        Training load set & - & 300-400 tasks/min \\
        Training seeds/episodes  (all method) & - & 10/3000 \\
        Micro-slot / macro-slot / ratio & $\Delta t$ / $T$ / $K$ & 50 ms / 1 s / 20 \\
        Task data size & $D_i$ & $\mathcal{U}(10, 40)$ Mbit \\
        Computation intensity & $C_i$ & $100$ cycles/bit \\
        Task max latency & $T_i^{\max}$ & $\mathcal{U}(2, 7.5)$ s \\
        \midrule
        \multicolumn{3}{c}{\textit{Training Hyperparameters}} \\
        \midrule
        Manager Actor learning rate & $\eta^M_a$ & $1 \times 10^{-4}$ \\
        Worker Actor learning rate & $\eta^W_a$ & $3 \times 10^{-4}$ \\
        Actor lr (PPO/DDPG/SAC) & - & $\{3,\,0.5,\,1\}\!\times\!10^{-4}$ \\
        Discount factor manager / worker & $\gamma$ & 0.99 / 0.95 \\
        Interface sizes & $N_{\max}$/$K_w$/$N_{\mathrm{cap}}$ & 10 / 10 / 5 \\
        Reward weights & $\rho_{\mathrm{succ}}$, $\rho_{\mathrm{fail}}$ & 2, 1 \\
        PPO clip / bid margin & $\epsilon$ / $\varepsilon$ & 0.2 / $10^{-3}$ \\
        \bottomrule
    \end{tabular}}
\end{table}

\begin{figure*}[t]
    \centering
    \begin{subfigure}{0.285\textwidth}
        \centering
        \includegraphics[width=\linewidth]{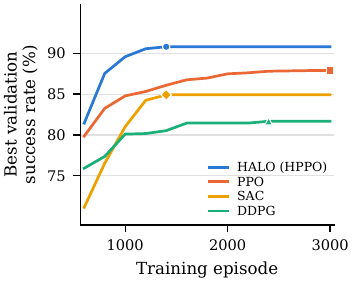}
        \caption{\small{Converged success rate (validation seeds, $\lambda=400$)}}
        \label{fig:convergence}
    \end{subfigure}
    \hfill
    \begin{subfigure}{0.285\textwidth}
        \centering
        \includegraphics[width=\linewidth]{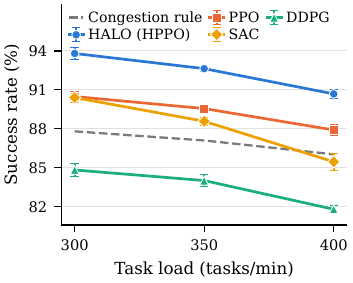}
        \caption{\small{Success rate (mean $\pm$ 95\% CI)}}
        \label{fig:baseline}
    \end{subfigure}
    \hfill
    \begin{subfigure}{0.285\textwidth}
        \centering
        \includegraphics[width=\linewidth]{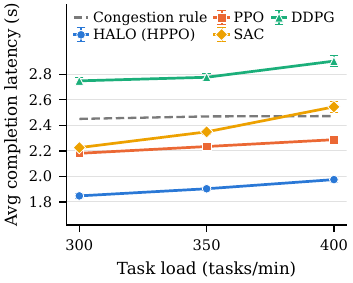}
        \caption{\small{Mean completion latency (mean $\pm$ 95\% CI)}}
        \label{fig:latency}
    \end{subfigure}

    \caption{\small{Performance across traffic loads (deterministic evaluation, 10 test traffic seeds per load in (b) and (c))}}
    \label{fig:three_results}
\end{figure*}


We compare the proposed HALO framework with three DRL-based approaches, namely PPO, DDPG, and SAC. We also consider a congestion rule-based non-learning baseline. In this, a task is processed locally when the device queue is empty. The task is offloaded to the HAPS if its size exceeds $35$\, Mbit, and to the UAV otherwise. Here, compute and channel resources are shared equally among active tasks. All methods use the same network configuration, task arrival process, resource constraints, and training budget to ensure a fair comparison. Each learning method is trained using 10 random seeds for 3000 episodes. The final model is selected using two separate validation seeds that are not used for training or testing.

For performance evaluation, we consider task arrival rates of
$\lambda \in \{300,350,400\}$ tasks/min. At each load, the trained policies are evaluated using the same 10 independent traffic seeds for all methods. The reported results show the mean performance together with the corresponding 95\% confidence interval. The main simulation and training parameters are summarized in Table~I.

\subsection{Convergence and Performance Comparison}

Fig.~3(a) shows the convergence behavior of the learning-based methods at the highest traffic load of 400 tasks/min. The curves represent the best validation success rate obtained during training. All four methods converge within the 3000-episode training budget. HALO reaches its best validation performance relatively early, at around 1400 episodes, while PPO continues improving until later stages of training. SAC becomes almost stable after approximately 1500 episodes, whereas DDPG shows noticeable variation across different training runs. Overall, HALO provides strong convergence performance with better consistency across training seeds.

Fig.~3(b) compares the task success rate under different traffic loads. HALO achieves success rates of 93.8\%, 92.6\%, and 90.7\% at 300, 350, and 400 tasks/min, respectively. As expected, the success rate decreases as the traffic load increases because more tasks compete for the available communication and computation resources. However, HALO consistently achieves the highest success rate at all three traffic loads.

Compared with PPO, HALO improves the success rate by approximately 3.7\%, 3.4\%, and 3.2\% at 300, 350, and 400 tasks/min, respectively. The corresponding improvements over DDPG are 10.6\%, 10.3\%, and 10.9\%, while the improvements over SAC are 3.8\%, 4.6\%, and 6.1\%. Averaged over all traffic loads, HALO provides relative gains of approximately 3.4\% over PPO, 10.6\% over DDPG, and 4.8\% over SAC. Moreover, HALO outperforms the learning-based baselines on all 10 shared test seeds at every evaluated traffic load. The congestion-based non-learning scheme achieves success rates of 87.8\%, 87.1\%, and 86.0\% at 300, 350, and 400 tasks/min, respectively. Nevertheless, HALO remains ahead of this scheme by 6.0, 5.5, and 4.7 percentage points. These results show that HALO is more effective at maintaining successful rate as the network becomes increasingly congested.

Fig.~3(c) compares the average task completion latency under different traffic loads. HALO achieves the lowest latency at every load, with average values of approximately 1.9 s, 1.95 s, and 2.0 s at 300, 350, and 400 tasks/min, respectively. In comparison, PPO requires approximately 2.2-2.3 s, DDPG requires approximately 2.8-2.9 s, and SAC increases from approximately 2.2 s to 2.6 s as the traffic load increases.

Relative to PPO, HALO reduces the average completion latency by approximately 15.3\%, 14.8\%, and 13.7\% at the three considered loads. Compared with DDPG, the latency reductions are approximately 32.8\%, 31.5\%, and 32.0\%, while reductions of 17.0\%, 19.0\%, and 22.4\% are achieved over SAC.

Overall, Fig.~3 demonstrates that HALO provides two important benefits simultaneously. First, more tasks are completed within their latency constraints. Second, the successfully completed tasks require less time to finish. The performance advantage remains consistent as the traffic load increases, indicating that the proposed hierarchical macro-micro resource management can more effectively handle congestion and dynamically allocate communication and computation resources.

\section{Conclusion}
In this paper, we proposed \textbf{HALO}, a hierarchical auction-assisted learning framework for delay-aware task offloading and resource scheduling in three-tier SAGIN. By combining a UAV-centric auction mechanism for task association with a two-timescale HPPO-based resource allocation strategy, HALO effectively coordinates discrete and continuous decisions under dynamic network conditions. The proposed macro-micro slot model enables fine-grained tracking of transmission and computation progress, improving latency awareness and system responsiveness. Simulation results demonstrate that HALO consistently outperforms conventional DRL baselines in both task success rate and latency, while maintaining robust performance under increasing traffic loads. These results highlight the effectiveness of hierarchical learning in managing complex SAGIN environments.  Future work will explore graph-structured state modeling for larger-scale SAGIN scenarios.

 \section*{Acknowledgment }\label{Section6}
 This work is supported in part by Natural Sciences and Engineering Research Council of Canada (NSERC) under the CREATE TRAVERSAL and NSERC DISCOVERY programs, and in part by the Ontario Research fund-Research Excellence (ORF-RE) program under project number RE12-024.\looseness=-1

\let\oldthebibliography\thebibliography
\renewcommand{\thebibliography}[1]{\oldthebibliography{#1}%
  \setlength{\itemsep}{0pt}\setlength{\parskip}{0pt}\setlength{\parsep}{0pt}}
{\bibliographystyle{IEEEtran}
\bibliography{HAPS}}

\end{document}